\documentclass[twocolumn,showpacs,preprintnumbers,amsmath,amssymb]{revtex4}

\usepackage{graphicx}% Include figure files
\usepackage{dcolumn}% Align table columns on decimal point
\usepackage{bm}% bold math

\begin{document}

\preprint{Physical Review B}

\title{Impurity-induced singlet breaking in SrCu$_2$(BO$_3$)$_2$}

\author{A.A. Aczel}
\author{G.J. MacDougall}
\author{J.A. Rodriguez}
\author{G.M. Luke}
\email{luke@mcmaster.ca}
\affiliation{Department of Physics and Astronomy, McMaster University, 1280 Main Street W., Hamilton, ON, Canada, L8S-4M1}
\author{P.L. Russo}
\thanks{Present address: TRIUMF, 4004 Wesbrook Mall, Vancouver, BC, Canada, V6T 2A3}
\affiliation{Department of Physics, Columbia University, 538 W. 120th St., New York, NY, USA, 10027}
\author{A.T. Savici}
\thanks{Present address: Condensed Matter Physics and Materials Science Department, Brookhaven National Lab, Upton, NY, USA, 11973}
\affiliation{Department of Physics, Columbia University, 538 W. 120th St., New York, NY, USA, 10027}
\author {Y.J. Uemura}
\affiliation{Department of Physics, Columbia University, 538 W. 120th St., New York, NY, USA, 10027}
\author{H.A. Dabkowska}
\affiliation{Brockhouse Institute for Materials Research, McMaster University, 1280 Main Street W., Hamilton, ON, Canada L8S-4M1}
\author{C.R. Wiebe}
\author{J.A. Janik}
\affiliation{Department of Physics, Florida State University, 211 Westcott Bldg., Tallahassee, FL, USA, 32306}
\author{H. Kageyama}
\affiliation{Department of Chemistry, Graduate School of Science, Kyoto University, Kyoto, 606-8502, Japan}

\date{\today}% It is always \today, today,
             % but any date may be explicitly specified

\begin{abstract}
We have performed $\mu$SR studies on single crystals of SrCu$_2$(BO$_3$)$_2$, a quasi-two-dimensional spin system with a spin singlet ground state. We observe two different muon sites which we associate with muons located adjacent to the two inequivalent O sites. One site, presumed to be located in the Cu-O-Cu superexchange path, exhibits a large increase in the frequency shift with decreasing temperature which is unaffected by the singlet formation, indicating that the muon has locally broken the singlet bond. We have also performed $\mu$SR on single crystals of SrMg$_{0.05}$Cu$_{1.95}$(BO$_3$)$_2$, Sr$_{0.96}$La$_{0.04}$Cu$_2$(BO$_3$)$_2$, and Sr$_{0.95}$Na$_{0.05}$Cu$_2$(BO$_3$)$_2$. We have found that the frequency shifts of these doped samples are equivalent and contain three branches at low temperatures. Two of these branches map on to the branches observed in the pure sample reasonably well and so we attribute the third branch to effect of the dopants. Specifically, this third branch represents the case when the muon sits at a site in the superexchange path lacking a corresponding singlet, due to it already being broken as a result of doping. This then leads to the conclusion that singlets are broken in this system when it is doped both in and out of the CuBO$_3$ planes.
\end{abstract}


\maketitle

\section{\label{sec:level1}Introduction}
Low-dimensional spin systems with spin-singlet ground states have attracted considerable study over the past 15 years. These systems are interesting for a number of reasons, including possible relevance to high temperature superconductivity and the existence of novel ground states. Most of the systems studied to date have been limited to quasi-one-dimension. These include spin ladder systems (e.g. SrCu$_2$O$_3$\cite{azuma_94}), spin-Peierls systems, (e.g. CuGeO$_3$\cite{hirota_94}) and Haldane gap systems (e.g. Y$_2$BaNiO$_5$\cite{darriet_93}). These materials all have a spin gap in the energy spectrum, which is associated with the spin singlet ground states.  Theoretical studies of spin ladders\cite{dagotto_92} have predicted d-wave superconductivity at low concentrations of hole doping in these systems. In fact, the mixed ladder-chain system Sr$_{14-x}$Ca$_x$Cu$_{24}$O$_{41}$ has been found to be a superconductor under pressure, specifically in the cases of x$=$13.6\cite{uehara_96} and x$=$11.5,\cite{nagata_98} with the
superconductivity thought to originate in the ladder layers. High-{\it T$_c$} cuprates have d-wave superconducting states and a partial (or pseudo) spin gap, which provides further evidence for strong similiarities between doped ladders and doped CuO$_2$ planes and motivates study of the effects of perturbing spin gap systems.

Two dimensional systems with spin singlet ground states are comparatively rare and often involve lattices where the active spins have a small number of nearest neighbours, such as CaV$_4$O$_9$,\cite{taniguchi} where the depleted square lattice consists of plaquettes of four spins, coupled to neighbouring plaquettes through one vertex. SrCu$_2$(BO$_3$)$_2$ is a second example of a quasi-two-dimensional spin system with a spin-singlet ground state. \cite{kageyama_99} The unit cell is tetragonal and the structure is characterized by layers consisting of Cu$^{2+}$, O$^{2-}$ and B$^{3+}$ running perpendicular to the c-axis. These layers are separated from each other by planes composed of Sr$^{2+}$ ions. All Cu$^{2+}$ sites have localized spin S = 1/2 moments and one nearest neighbour with which they form dimeric units, while each of these units is surrounded by four orthogonal dimers. This configuration of spins corresponds to the two-dimensional model that Shastry and Sutherland first introduced in 1981.\cite{shastry_81} This model is topologically equivalent to a two-dimensional Heisenberg model with nearest-neighbour coupling J and next-nearest-neighbour coupling J$^\prime$. With decreasing J$^\prime$/J, a phase transition from a four-sublattice Neel state to a gapped spin singlet state was shown to take place at a critical value of J$^\prime$/J~$\sim$~0.7. The ratio of J$^\prime$/J in SrCu$_2$(BO$_3$)$_2$ is about 0.68, a little bit smaller than the critical value.\cite{liu_05} High resolution inelastic neutron measurements\cite{gaulin_04} have identified one, two, and three triplet excitations. These measurements clearly show the appearance of the energy gap in the spectrum of excitations with decreasing temperature, and from them the energy gap is evaluated to be $\sim$~34 K. The one and two triplet excitations both display a very large drop-off in inelastic intensity with increasing temperature. This temperature dependence is generally not seen in a system undergoing a phase transition, and no evidence of symmetry breaking has been associated with the appearance of the spin singlet ground state. Another interesting feature of this system is the magnetization plateaus that are observed at 1/8, 1/4,\cite{kageyama_99} and 1/3\cite{onizuka_00} of the saturation magnetization in high magnetic fields ($>$~20 T). This is the first 2D system where these plateaus have been observed.

While superconductivity has been predicted and observed in some quasi-one-dimensional systems with a spin singlet ground state, it has also been predicted in the doped Shastry-Sutherland model.\cite{shastry_02, chung_04, kimura_04} Liu et. al have recently investigated a series of polycrystalline SrCu$_2$(BO$_3$)$_2$ samples, with dopants added both in\cite{liu_06} and out\cite{liu_05} of the CuBO$_3$ planes. They found that the spin gap behaviour was strongly affected by doping, but did not find any evidence for superconductivity. The resistivity of the pure sample at room temperature was found to be 41.73 M$\Omega$m,\cite{liu_05} and the resistivity of all doped samples was found to be only one or two orders of magnitude smaller. This implies that all their samples were insulators and that the carriers introduced as a result of doping were not very mobile, at least at room temperature. However, the theoretical predictions discussed above make it important to better understand Shastry-Sutherland systems such as SrCu$_2$(BO$_3$)$_2$ and their response to perturbations.

Muon spin relaxation measurements of a spin singlet state give a signal characteristic of a non-magnetic state. Zero field measurements\cite{kk97}
of the spin-Peierls system CuGeO$_3$ gave essentially a temperature-independent signal, with weak relaxation being ascribed to isolated defects associated with impurities. In the case of SrCu$_2$(BO$_3$)$_2$, Fukaya {\em et al.}\cite{fukaya_03} reported zero field (ZF) and longitudinal field (LF) $\mu$SR measurements of ceramic specimens. In contrast to their expectations of seeing no spin relaxation associated with this state, the authors reported the onset of a large, roughly constant relaxation for temperatures of 3 K and below. They attributed this relaxation to dilute defects or to $\mu$SR detecting weak intrinsic magnetism that other probes could not see. The authors also reported that dynamic fluctuations persisted in the system down to base temperature (20 mK) and that these fluctuations likely have a quantum origin.

Lappas {\em et al.} also investigated ceramic samples of SrCu$_2$(BO$_3$)$_2$ using the $\mu$SR technique.\cite{lappas_03} The authors reported the same anomalous relaxation in ZF as observed by Fukaya {\em et al.} using a different batch of samples. They attributed this relaxation to an unusual spin freezing process. The authors went on to perform 0.6 T frequency shift measurements on their samples. They concluded that implanted muons may liberate spin density in the spin gap regime that spin-freezes at very low temperatures.

\section{\label{sec:level2}Experimental Details}
We grew single crystals of SrCu$_2$(BO$_3$)$_2$, SrMg$_{0.05}$Cu$_{1.95}$(BO$_3$)$_2$, Sr$_{0.96}$La$_{0.04}$Cu$_2$(BO$_3$)$_2$, and Sr$_{0.95}$Na$_{0.05}$Cu$_2$(BO$_3$)$_2$ in floating zone image furnaces at Tokyo\cite{kageyama_99_2} and McMaster University\cite{dabkowska_06}
following slightly different procedures. Neutron scattering measurements have been reported by Haravifard et al.\cite{haravifard_06} on the system SrMg$_{0.05}$Cu$_{1.95}$(BO$_3$)$_2$. Aside from the original crystal growth article\cite{dabkowska_06}, this seems to be the only paper describing doped single crystals of this system as they have proven to be quite difficult to grow. 

\begin{figure}
\begin{center}
\scalebox{0.3}{\includegraphics[angle=90]{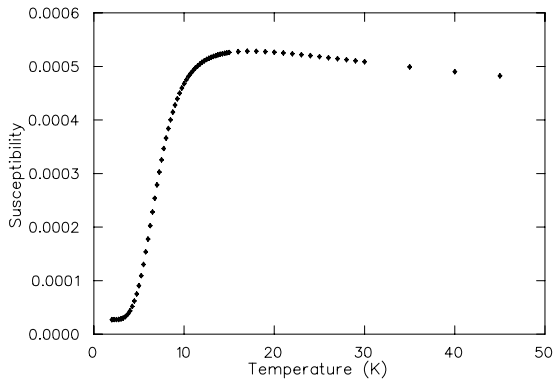}}
\caption{\label{SCBO_susceptibility} DC volume susceptibility of single crystal SrCu$_2$(BO$_3$)$_2$, measured in an applied magnetic
field H~=~5~T~$\parallel~\hat{c}$.}
\end{center}
\end{figure}

DC susceptibility measurements of a McMaster crystal of SrCu$_2$(BO$_3$)$_2$ in a field of 5~T parallel to the c-axis are shown in Fig.~\ref{SCBO_susceptibility}; results for the Tokyo crystals are essentially identical. There is a sharp decrease in the susceptibility below 15 K which provides evidence for the spin singlet ground state. A small upturn is seen in the susceptibility curve below 4 K, which likely arises due to magnetic impurities and/or defects of Cu$^{2+}$ ions in SrCu$_2$(BO$_3$)$_2$. The susceptibility data below 7.5 K was fit to the following functional form: 
\begin{equation}
\chi=C/T+Be^{-\Delta/T}+\chi_0
\end{equation}
The first Curie term is due to the contribution of magnetic impurities, the second term is to account for the spin gap, and the third temperature-independent term accounts for Van-Vleck paramagnetism and core diamagnetism. The spin gap $\Delta$ was found to be $\sim$ 20.2(7) K in our field of 5 T, which is in line with other reported estimates. If we ascribe the Curie term to free Cu spins, then this concentration would be only 0.12(4)~\%. 

We performed 5 T transverse field (TF) $\mu$SR measurements on the pure and doped SrCu$_2$(BO$_3$)$_2$ samples to investigate the presumed non-magnetic spin singlet ground states in these systems. These measurements were conducted on the M15 surface muon channel at TRIUMF in Vancouver, Canada, using a helium gas flow cryostat in the temperature range 2 K $<$ T $<$ 150 K. The samples were mounted in a low-background spectrometer with the c-axis parallel to the incoming muon momentum (and the external field). In this high field experiment, the measurements of the muon precession frequency signal in our samples and a reference material (CaCO$_3$) were taken simultaneously to allow for a precise determination of the frequency shift.

In addition, we performed ZF-$\mu$SR and low TF-$\mu$SR measurements (H $<$ 1 T) on the pure sample. After observing the anomalous relaxation previously seen by Fukaya {\em et al.}\cite{fukaya_03} in the powder sample, we performed LF-$\mu$SR on our single crystal to characterize the nature of this relaxation. For these low temperature measurements (between 20 mK and 10 K), we used an Oxford Instruments dilution refrigerator on the M15 beam line. The sample was mounted on a silver sample holder, with the c-axis parallel to the incoming muon beam, as in the case of the high TF measurements.

\section{\label{sec:level3}Discussion and Analysis}

\begin{figure}
\begin{center}
\scalebox{0.3}{\includegraphics{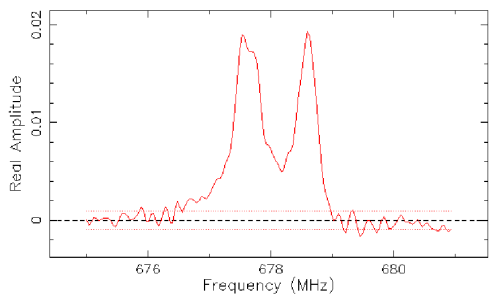}}
\caption{\label{pure_FFT} (Color online) Fast Fourier transform of muon precession signal for SrCu$_2$(BO$_3$)$_2$ measured in an applied magnetic field H~=~5~T~$\parallel ~\hat{c}$ at 20 K. The two peaks correspond to two magnetically inequivalent muon sites.}
\end{center}
\end{figure}

Implanted positive muons reside at sites which are minima in the electrostatic potential. Generally in oxides, muons are hydrogen-bonded to O$^{2-}$ ions, roughly 1~\AA\ away. Fig. \ref{pure_FFT} shows a fast Fourier transform of the transverse field muon precession signal measured in an applied magnetic field H = 5 T $\parallel\hat{c}$ at T = 20 K. Two peaks are clearly visible which correspond to two magnetically-inequivalent muon sites in SrCu$_2$(BO$_3$)$_2$; we associate these with the two magnetically-inequivalent oxygen sites in the system. To take these two muon sites into account, this 5 T TF data was fit to the sum of two exponentially-damped cosine functions. Our exact fitting function is given by:
\begin{equation}
P(t)=A_1\exp^{-\lambda_1t}\cos(\omega_1t+\phi_1)+A_2\exp^{-\lambda_2t}\cos(\omega_2t+\phi_2)
\end{equation}
where $\omega_1$ and $\omega_2$ are the average muon precession frequencies, $\lambda_1$ and $\lambda_2$ are the muon relaxation rates for the two sites, and $A_1$ and $A_2$ represent the relative occupancy of the two unique muon sites. $A_1$ and $A_2$ are both temperature independent and were found to have a ratio of roughly 3:2. The ratio of the two inequivalent O sites in the unit cell is 2:1, with fewer O sites lying in the Cu-O-Cu superexchange path. However, the probability of occupying a specific site depends not just on the relative number of such sites, but also on the exact energies of those sites. Generally, there should only be one true stable muon site, but a metastable site could be long-lived on the time range of our experiment, leading to the site-occupation ratio that we observe here.

\begin{figure*}
\begin{center}
\scalebox{0.55}{\includegraphics[angle=90]{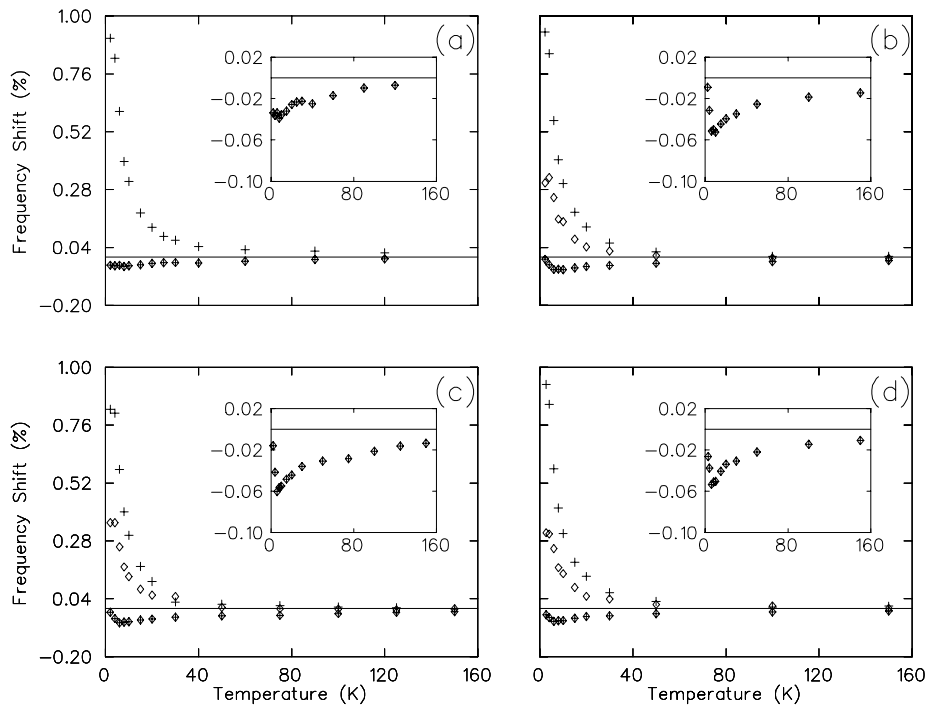}}
\caption{\label{5T_knightshift} Frequency shifts vs. temperature measured in an applied field of 5 T for: (a) SrCu$_2$(BO$_3$)$_2$, (b) Sr$_{0.96}$La$_{0.04}$Cu$_2$(BO$_3$)$_2$, (c) SrMg$_{0.05}$Cu$_{1.95}$(BO$_3$)$_2$, and (d) Sr$_{0.95}$Na$_{0.05}$Cu$_2$(BO$_3$)$_2$. The insets show the temp. dependence of the frequency shift for the muon site outside of the superexchange path.}
\end{center}
\end{figure*}

The frequency shift is the fractional difference between the applied field and the local field at the muon site, and it is proportional to the local spin susceptibility. We see that for one of the muon sites the frequency shift roughly follows a Curie-Weiss temperature dependence for T $>$ 4 K (represented by the crosses in Fig.~\ref{5T_knightshift}a), with the precise value of the Curie constant depending on the temperature range being fit. This is in contrast to $^{11}$B-NMR frequency shift measurements.\cite{kodama_04} In that case, the authors applied a field of 7 T perpendicular to the [1 -1 0] direction and observed a decrease in the frequency shift below the singlet formation temperature of 15 K. The lack of a similar decrease in our $\mu^+$ frequency shift for this site indicates the presence of a larger local field than expected in the singlet regime. The fundamental difference between the two experiments is that in our case there is a muon present in the system, which means the larger local field is likely the result of a muon perturbation effect. More specifically, it seems that the muon has locally broken a spin singlet bond and created at least two quasi-free spins. These spins are then polarized by the applied field which leads to our observed frequency shift. As this appears to be a strong effect, it then seems reasonable to associate this muon site with the oxygen site in the Cu-O-Cu superexchange path.

The shift at the other muon site is considerably smaller and negative (see inset of Fig.~\ref{5T_knightshift}a). In this case, the frequency shift follows the temperature-dependence of the bulk susceptibility reasonably well, indicating the lack of any strong muon perturbation. For this reason, this site likely corresponds to the O$^{2-}$ ion out of the Cu-O-Cu superexchange path. 

Muon perturbation effects have been observed in other materials. In the system PrNi$_5$, it has been shown that an implanted muon locally alters the crystal field levels.\cite{feyerherm_95} Muon-induced singlet-breaking, similar to what is being described in this work, has been observed in the spin-1/2 ladder system KCuCl$_3$.\cite{andreica_00} An implanted muon has also been reported to perturb a link in the AF chain of the quasi 1-D systems KCuF$_3$\cite{chakhalian_03} and dichlorobis (pyridine) copper (II)\cite{chakhalian_03_2}, where the observed local spin susceptibility agreed well with theoretical calculations.\cite{affleck_95, affleck_92} Our measurement of the local susceptibility at two different sites in SrCu$_2$(BO$_3$)$_2$ should serve to stimulate similiar calculations for the two-dimensional Shastry-Sutherland system. 

Motivated by the discovery of the two muon sites in SrCu$_2$(BO$_3$)$_2$, we fit the ZF-$\mu$SR data to the sum of two terms, each taken to be the product of an exponential and a static Gaussian Kubo-Toyabe relaxation function. The Gaussian Kubo-Toyabe part represents relaxation due to nuclear moments that are oriented randomly in ZF. This is generally a temperature-independent contribution to the relaxation and this was assumed in this case by fixing the Gaussian Kubo-Toyabe relaxation rates. The ratio of the asymmetries of the two sites was fixed to the value obtained when fitting the high TF data. Fig. 4 shows the exponential relaxation of one site is roughly independent of temperature for T $>$ 4 K, then rises sharply at 4 K and saturates for temperatures of 3 K and below. The other site has a very small, almost temperature-independent relaxation. The large increase and saturation of the exponential relaxation for the one muon site at low temperatures may imply slowing down or freezing of the spins liberated by the muon perturbation effect as previously suggested by Lappas {\em et al.}.\cite{lappas_03}

\begin{figure}
\begin{center}
\scalebox{0.3}{\includegraphics[angle=90]{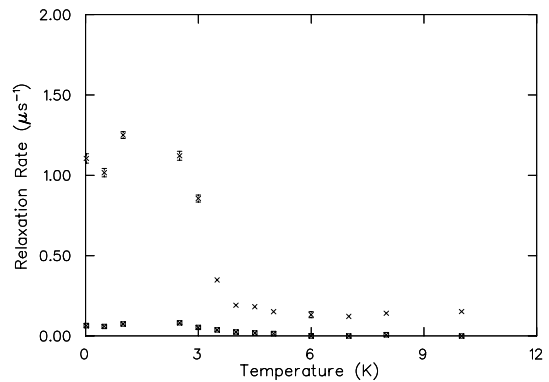}}
\caption{\label{ZF_rlx} Exponential relaxation rates vs. temperature for zero applied field in SrCu$_2$(BO$_3$)$_2$.}
\end{center}
\end{figure}

\begin{figure}
\begin{center}
\scalebox{0.3}{\includegraphics{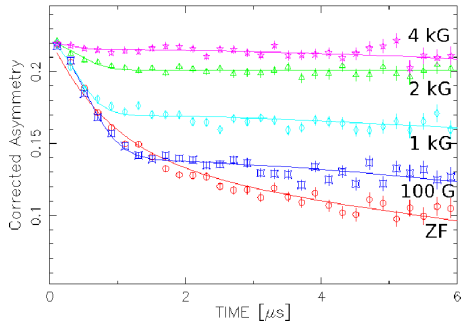}}
\caption{\label{LF_lineshapes} (Color online) LF-$\mu$SR line shapes for selected applied fields at 20 mK. There is significant relaxation even
in the case of 2 kG. The lines are a guide for the eye.}
\end{center}
\end{figure}

\begin{figure}
\begin{center}
\scalebox{0.3}{\includegraphics{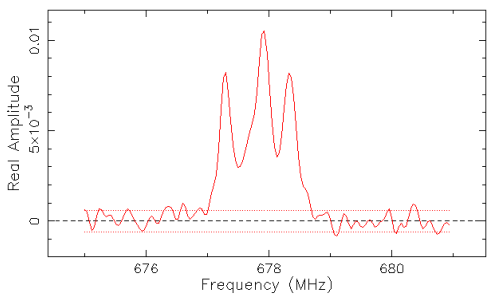}}
\caption{\label{Mg_FFT} (Color online) Fast Fourier transform of precession signal for SrMg$_{0.05}$Cu$_{1.95}$(BO$_3$)$_2$ measured in an applied magnetic field H~=~5~T~$\parallel\hat{c}$ at 20 K. There are now three peaks instead of two; the third peak is due to the Mg$^{2+}$ ions.}
\end{center}
\end{figure}

\begin{table*}
\caption{\label{table} Estimated spin gaps, Curie constants, and fraction of free spin 1/2-Cu's by fitting susceptibility data with H = 1 kG $\parallel$ ab plane.}
\begin{ruledtabular}
\begin{tabular}{ccccc}
 Sample&Spin Gap (K)&Curie Constant (K)&Estimated fraction of free Cu spins (\%)\\ \hline
 SrCu$_2$(BO$_3$)$_2$&27.8(8)&3.4$\times10^{-5}$(8)&0.0585(1) \\
 SrMg$_{0.05}$Cu$_{1.95}$(BO$_3$)$_2$&20.4(6)&5.5$\times10^{-4}$(1)&0.95(2) \\
 Sr$_{0.96}$La$_{0.04}$Cu$_2$(BO$_3$)$_2$&23.2(6)&3.09$\times10^{-4}$(1)&0.53(2) \\
 Sr$_{0.95}$Na$_{0.05}$Cu$_2$(BO$_3$)$_2$&24.4(5)&1.67$\times10^{-4}$(1)&0.29(2) \\
\end{tabular}
\end{ruledtabular}
\end{table*}

To see if this was a suitable description, it was necessary to examine longitudinal field (LF) $\mu$SR data taken below the saturation temperature, specifically at 20 mK in this case (Fig.~\ref{LF_lineshapes}). Values of $\sim$ 5 G and 80 G were estimated for the local fields at the two muon sites from the ZF time spectrum at T $=$ 20 mK using the relation B $\sim \lambda/\gamma_{\mu}$ ($\gamma_{\mu}=135.5$ MHz/T) with $\lambda$ representing the muon relaxation rate as before. Assuming these local fields are static, a comparable LF (up to one order of magnitude larger) should be enough to completely decouple them. However, we are still seeing significant relaxation even in the case of LF $=$ 2 kG. This implies that the spins are dynamic in the low temperature regime and their behaviour cannot be described by spin freezing. These spin fluctuations are fast for T~$>$~4 K and then slow down substantially below T~$\sim$~4 K. The saturation of the relaxation at the low temperatures suggests that these fluctuations are not thermal but instead have a quantum origin. This is in agreement with what Fukaya {\em et al.} reported in the ceramic samples.\cite{fukaya_03} If the LF spectra are examined closely at early times, it becomes clear that they have a distinct Gaussian nature. In general, dynamic spin fluctuations cannot be described by a Gaussian lineshape. This appears to be one of the rare cases of undecouplable Gaussian behaviour in $\mu$SR, which is still poorly understood. Such behaviour has been previously observed in frustrated and/or low-dimensional systems.\cite{uemura_94, fukaya_03_2}

Fig. \ref{Mg_FFT} shows a fast Fourier transform of the transverse field muon precession signal measured in an applied magnetic field H = 5 T $\parallel\hat{c}$ at T = 20 K for SrMg$_{0.05}$Cu$_{1.95}$(BO$_3$)$_2$. Similar FFTs were obtained for the La and Na-doped samples. There are clearly three frequency peaks in this case as opposed to the two that were seen in the FFT of the pure sample. For this reason, the data was fit to the same functional form as given by Eq. (2), but three exponentially-damped cosine terms were included instead of two. The frequency shifts calculated from this fit are given in Fig.~\ref{5T_knightshift}b-d; the shifts for all three doped samples examined appear to be essentially equivalent. The upper and lower branches are remarkably similar to the two branches in the pure case, but the middle branch is new and so it seems is associated with the dopants. 

In one situation, we are substituting non-magnetic Mg in for the Cu sites. It is well-known that this leads to the direct breaking of spin singlets, as some Cu spins no longer have a nearest neighbour Cu with which to form a dimer. In the other two cases, we are substituting La or Na in for the Sr sites. These are both forms of charge doping, with each La introducing one electron into our system and each Na introducing one hole. Each electron or hole introduced by doping is expected to break a singlet pair and make a localized spin-1/2. Resistivity measurements performed by Liu {\em et al.}\cite{liu_05} on La and Na-doped polycrystalline samples verify the localized nature of these carriers. The breaking of singlets by the different dopants is supported by DC susceptibility measurements performed on our single crystals with an applied field H = 1 kG $\parallel$ ab plane.\cite{dabkowska_06} All three doped samples had much larger Curie tails than the pure sample, indicating the presence of many more free spin-1/2~Cu's in the singlet regime. We have calculated the spin gap, Curie constant and the resulting fraction of free spins by fitting this data to Eq. (1); the results are shown in Table I. In all the doped cases, the spin gap has been supressed. Also, the fraction of free spins is at least five times as great as in the pure case, but still lower than what we expect considering the concentration of our starting materials. This is likely due to difficulties involved in doping this system and suggests that our doping concentrations should be viewed as nominal values. 

Our $\mu$SR results complement the susceptibility measurements by providing microscopic evidence that singlets are broken in this system when it is doped both in and out of the CuBO$_3$ planes. It appears there are three distinct possibilities for muons in the doped samples. The muon can either stop at the sites in or out of the superexchange path as before. If the muon stops at the site in the superexchange path, it will break the singlet and create two quasi-free spins in close proximity. However, in some cases it seems the singlet will already be broken due to the dopants and then the muon will have no singlet to break. In this case, the muon will only be in close proximity to one quasi-free spin and so a smaller frequency shift will be measured with decreasing temperature, giving us the middle branch. In fact, Figure \ref{5T_knightshift}b-d shows that this middle branch gives about half the shift of the upper branch. 

\section{\label{sec:level4}Conclusion}
We determined there are two different muon sites in SrCu$_2$(BO$_3$)$_2$, which we associate with muons located adjacent to the two inequivalent O sites in the system. One site exhibits a large increase in the frequency shift with decreasing temperature that is unaffected by singlet formation, indicating that the muon sits in the Cu-O-Cu superexchange path and has broken the corresponding dimer. The frequency shift of the other site is considerably smaller and negative. It roughly scales with the DC bulk susceptibility, indicating the lack of a perturbation effect in this case. 

We also observe a large relaxation increase for the site in the superexchange path in the singlet regime, which is in contrast to the weak, practically temperature-independent relaxation expected in such a state. Analysis of LF-$\mu$SR data has shown that this relaxation is likely due to slow but dynamic fluctuations of the liberated spins that persist down to our base temperature of 20 mK.

Finally, all three doped samples studied have remarkably similar frequency shifts. In each case, there are now three branches at low temperatures instead of two as observed in the pure sample. The upper and lower branches map on to the two branches in the pure sample very well, and so the third branch is attributed to the dopants. Specifically, this third branch represents the case when the muon sits at a site in the superexchange path lacking a corresponding singlet due to it already being broken as a result of doping.  Our results demonstrate that electron or hole doping of SrCu$_2$(BO$_3$)$_2$ by substituting on the Sr site gives the same result as substitution of Cu by non-magnetic Mg: namely, the breaking of a localized singlet and the liberation of a single Cu spin.  We find that the number of free spins is significantly less than expected from the nominal dopant concentration and that reaching doping levels sufficient for metallic (and superconducting) behavior may be difficult in this system.

\begin{acknowledgments}
We appreciate the hospitality of the TRIUMF Center for Molecular and Materials Science where the $\mu$SR experiments were performed. 

A.~A.~Aczel and G.~J.~MacDougall are supported by NSERC CGS Fellowships. Research at McMaster University is supported by NSERC and the Canadian Institute for Advanced Research Quantum Materials Program. Research at Columbia University is supported by NSF DMR-05-02706, NSF DMR-01-02752, and CHE-01-11752. 

\end{acknowledgments}

\end{document}